# What Personal Information Improves LLM-Based Next-Location Prediction?

Xin Wang [a], Páraic Carroll [a], Kerry Nice [a], Sachith Seneviratne [a], Li Zhang [b*]

[a] Transport, Health and Urban Systems Research Lab, Faculty of Architecture, Building and Planning, The University of Melbourne, Melbourne, Australia

[b] School of Architecture and Urban Planning, Shenzhen University, Shenzhen, China

## Abstract

Large language models (LLMs) are increasingly used for individual next-location prediction, with personal information easily added to prompts alongside mobility history. Yet the incremental predictive value of such information remains unclear. Using linked sociodemographic records and mobility traces from 5,000 Shenzhen residents, this study separates model responsiveness from predictive value. GPT-5 is the primary model, with GPT-5.5 and Claude Opus 4.6 used for replication. In 1,000 paired prediction instances, models rank 100 candidate destinations with and without age, gender, occupation and income while all other inputs are held fixed. Behavioural history raises top-1 accuracy from 5.6% to 18.5% as prior history increases from zero to six days. By contrast, sociodemographic attributes produce no detectable overall gain, although replacing correct attributes with those of another person reduces accuracy by 5.4 percentage points. Candidate construction also matters, removing distance raises accuracy by 7.7 points under proximity sampling but lowers it by 22.3 points under popularity sampling, with the reversal reproduced across all three LLMs. These findings identify behavioural history as the clearest source of incremental value and show that personal information should be evaluated under matched, explicitly specified conditions before its privacy and governance costs are justified.

# 1. Introduction

Individual next-location prediction underpins applications from transport demand management to epidemic response and location-based services, and large language models have recently been proposed as general-purpose predictors for this task (Wang et al., 2023; Beneduce et al., 2025; Feng et al., 2025; Tang et al., 2025; Yang et al., 2025). Given a person's recent stay sequence rendered as text, a large language model (LLM) is asked to name or rank the place that person will visit next. The appeal is that LLMs bring world knowledge and zero-shot generalisation to a problem where conventional sequence models require city-specific training data (Feng et al., 2018; Yang et al., 2020), and reported accuracies have been competitive with supervised baselines (Beneduce et al., 2025). Because LLMs accept natural-language input, personal characteristics can be readily incorporated alongside mobility history. This flexibility raises a more fundamental question than model choice: which personal information actually improves individual-level prediction, and how should its contribution be evaluated?

Existing evidence on the role of personal information in mobility modelling comes from largely separate strands of research. In studies that use LLMs to generate or simulate travel, sociodemographic conditioning is standard and often treated as a core component of the modelled persona. Persona-driven frameworks supply attributes such as age, gender, occupation and income before asking the model to produce a travel diary or survey response (Li et al., 2024; Tzachristas et al., 2026), following the broader generative-agent practice of grounding simulated behaviour in a described identity (Park et al., 2023). This practice has a plausible empirical basis at the population level, as national travel surveys consistently document systematic differences across sociodemographic groups in how far, how often and by what mode people travel (Hudecek et al., 2026). These regularities make sociodemographic attributes reasonable inputs when the aim is to generate behaviour that reflects differences between population groups. By contrast, existing LLM-based next-location predictors have been developed almost entirely from trajectory data, with sociodemographic attributes absent from their inputs. For example, LLM-Mob is evaluated on Geolife and Foursquare-NYC (Wang et al., 2023), while AgentMove uses Foursquare and ISP mobility traces (Feng et al., 2025); these datasets record where people move but do not link those trajectories to individual-level sociodemographic records. Canonical prediction prompts therefore centre on historical and contextual visits, while attempts to incorporate richer individual profiles derive them from observed mobility behaviour rather than external sociodemographic information (Feng et al., 2025). The available evidence linking attributes and mobility also runs mainly in the reverse direction. Attributes can be recovered from observed movement (Blumenstock et al., 2015; Uğurel et al., 2026), but this does not establish that supplying them improves prediction of where an individual goes next. Therefore, it remains unclear whether sociodemographic attributes provide incremental predictive value beyond an individual's observed mobility history, and more broadly which forms of personal information meaningfully improve LLM-based next-location prediction.

Four lines of work come closer without closing the gap. First, Solomon et al. (2021) train LSTM stay-point predictors on GPS trajectories and report how accuracy differs across demographic groups, which identifies who is easy to predict; it does not measure what the label adds. Second, persona studies do supply attributes, but they sample synthetic individuals from census marginals and ask whether the aggregate distribution of simulated behaviour matches survey statistics (Tzachristas et al., 2026); because attribute-behaviour coupling in synthetic data is imposed by the generator, such designs cannot recover the marginal value of a real person's real attributes. Third, bias audits provide real demographic labels but evaluate their effects without conditioning on an individual's observed mobility history. Wu and Wang (2024), for example, show that supplying race and gender can substantially shift predicted destinations, including the types of places associated with different demographic groups. This demonstrates that demographic cues can influence model predictions, but not whether they improve prediction when added to a person's own mobility history. A fourth line of research incorporates user profiles directly into prediction, but these profiles are derived from the same behavioural history used to predict future visits, as in ZeroPOIRec (Kim et al., 2025). Their apparent predictive contribution therefore does not show whether independently observed sociodemographic attributes provide additional information beyond mobility history. Across all four designs, that incremental contribution of personal information remains untested.

Resolving this question could inform how LLM-based mobility prediction systems are designed and evaluated. Supplying sociodemographic attributes to such systems requires access to and maintenance of individual-level personal data, bringing associated governance and data-protection obligations. It may also expose the model more directly to demographic cues that can shape or amplify stereotyped predictions (Wu & Wang, 2024). In addition, extra attribute information increases prompt complexity without necessarily adding task-relevant signal and may therefore dilute the predictive value of more directly informative mobility history. These considerations make incremental predictive value the key criterion for assessing whether sociodemographic attributes are useful for next-location prediction. This study therefore links sociodemographic records to mobility trajectories for 5,000 Shenzhen residents and constructs a closed-set next-location benchmark to compare the predictive value of behavioural history and sociodemographic information. The central question is whether sociodemographic attributes provide additional predictive value once an individual's own mobility history is already observed. Figure 1 summarises the design. Each instance is evaluated with mobility history alone and with the same history plus the person's sociodemographic attributes, and in a matched subset with attributes taken from another person, while holding the candidate set, target, model and all other prompt content constant. The paired design estimates the incremental predictive value of sociodemographic information across different amounts of mobility history. Complementary analyses then test whether the results generalise across models and prediction times, distinguish model responsiveness from improvements in predictive accuracy, and examine how candidate construction affects both the measured value of prompt information and the relative performance of competing predictors.

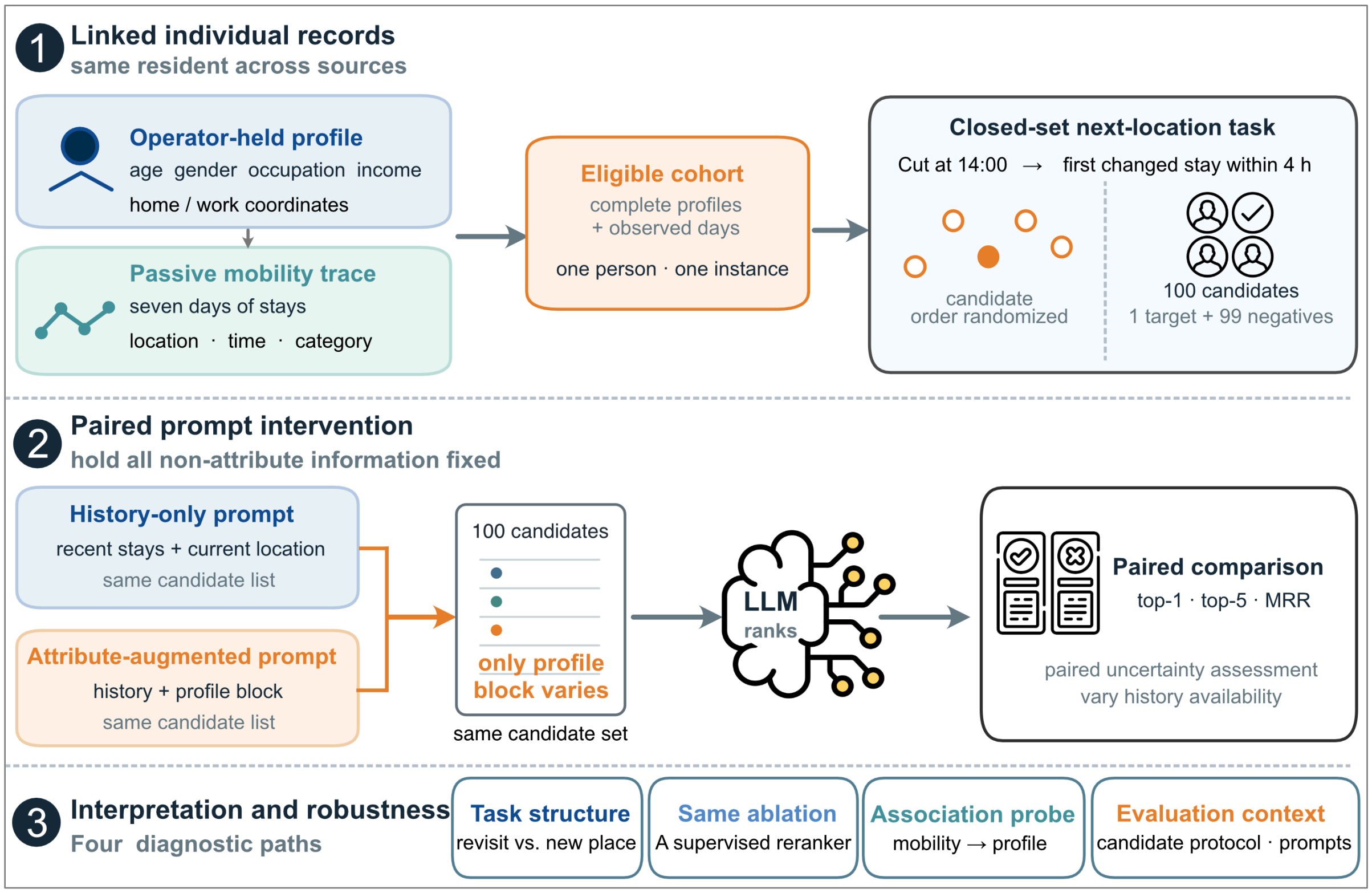

***Figure 1. Study framework**. Linked profiles and mobility traces form one closed-set instance per resident. The same candidate set is ranked with history-only and history-plus-attributes prompts, holding all other inputs fixed.*

*Further analyses examine task structure, a supervised predictor, demographic signal in trajectories and the candidate protocol.*

This study makes three contributions. Conceptually, it clarifies which forms of personal information add predictive value to LLM-based next-location prediction. The results distinguish the value of an individual's behavioural history from that of sociodemographic attributes and, in doing so, separate information that changes model outputs from information that improves prediction. Methodologically, the study links observed sociodemographic records with individual mobility trajectories, allowing the contribution of demographic conditioning to be tested directly while holding mobility history and other prediction inputs constant. The same linked data also make it possible to examine the relationship in both directions, comparing how much mobility behaviour reveals about sociodemographic characteristics with how much those characteristics contribute to predicting subsequent mobility. Practically, the study provides evidence on whether potentially sensitive sociodemographic information contributes enough predictive value to justify its inclusion in individual mobility prediction systems.

The remainder of the paper is organised as follows. Section 2 describes the linked dataset, the benchmark construction and the paired experimental design. Section 3 reports, in turn, the predictive value of behavioural history, the response to and benefit of demographic attributes, how far the absence of a gain extends across predictors and prediction times, and the influence of candidate construction, followed by an auxiliary analysis of demographic signal in trajectories. Section 4 discusses implications for persona-conditioned mobility modelling and for evaluation practice, and states the limitations that bound our conclusions. Section 5 concludes.

# 2. Data and Study Design

## 2.1 Linked individual attributes and mobility trajectories

This study uses a person-level dataset linking sociodemographic attributes to passively sensed mobility trajectories for residents of Shenzhen, China. The observation window covers seven consecutive days from 12 to 18 March 2024, comprising five weekdays and two weekend days. The mobility data contain device-level location observations with coordinates and timestamps, with approximately 72.6 million records per day across Shenzhen. The associated attribute records include age band, gender, occupation, income, and home and workplace coordinates. The two data sources were linked through a common device identifier, and the consistency of this linkage was evaluated empirically. For 71.4% of individuals, the community most frequently visited in the observed trajectory matched the home community identified from the corresponding attribute record. This proportion fell to 0.2% when device identifiers were randomly permuted across attribute records. Similarly, the median distance between the trajectory centroid and the recorded home location was 1.85 km for the observed linkage, compared with 19.99 km under random permutation. These comparisons provide independent evidence that the linked mobility and sociodemographic records correspond to the same individual.

Eligibility criteria were applied to reduce incomplete observation and occupational confounding. A person-day was considered valid if it contained at least five location observations within the Shenzhen administrative boundary and the interval between the first and last observation was at least six hours. Individuals were retained if at least six of the seven days met this criterion, all required attributes were available, and occupation was not recorded as food delivery courier or taxi driver. Couriers and taxi drivers were excluded because their mobility patterns are strongly structured by platform dispatching and may therefore reflect algorithmically assigned routing in addition to individual travel behaviour. Of the 4,181,584 devices observed at least once during the study period, 2,310,157 met the valid-day criterion and 666,390 remained after requiring complete attributes and excluding couriers. A seeded random sample of 5,000 individuals was drawn from this eligible pool for analysis. The sample closely reproduced the pool distributions of gender and age band, with deviations of no more than 0.8 percentage points for any category, indicating minimal compositional distortion from the sampling step.

Raw location observations were converted into stays, semantic activity labels, and administrative spatial units using a fixed preprocessing pipeline. Consecutive observations within 150 m of a cluster anchor were merged into a single stay. Each stay was then matched to the nearest point of interest within 50 m, producing a match rate of 90.8% and a median matching distance of 9.7 m. Spatial assignment to one of 648 administrative communities was performed using a point-in-polygon operation. Direct containment accounted for 98.1% of stays, with the remaining observations assigned to the nearest community polygon. The resulting dataset contained 330,342 stays, corresponding to an average of 9.48 stays per observed person-day. Coordinate reference systems were reconciled before spatial matching, and observations outside the Shenzhen administrative boundary were excluded before subsequent analysis.

The anonymized location-based service data were provided by Moxing Beijing for non-commercial academic research. The data were originally collected through smartphone applications with users' consent and supplied to the research team in de-identified form. Individual users were represented by anonymized identifiers, with no names, phone numbers, email addresses, or other information linking these identifiers to identifiable individuals. No participants were contacted or directly involved in the study. For the LLM experiments, the data were further minimized before being submitted to the models. Prompts contained neither anonymized device identifiers nor geographic coordinates, home locations, or workplace locations. Locations were instead represented using pseudonymous labels together with mobility-related information such as semantic categories, relative distances, visit times and, where applicable, coarse sociodemographic attributes. This additional preprocessing ensured that the information supplied to the LLMs was limited to that required for the prediction task. API-submitted data were not used by the model providers for model training under the applicable service terms.

## 2.2 Prediction task and benchmark design

Next-location prediction is formulated as a closed-set ranking task in which each model ranks 100 candidate locations containing one observed continuation and 99 alternatives. A 100-item candidate set follows established sampled-ranking practice in next-POI and sequential recommendation evaluation (Lou & Cui, 2026; Yu et al., 2026) and provides a sufficiently large, computationally tractable choice space for controlled model comparison. The candidate set is an evaluation construct and is not intended to represent the traveller's behavioural choice set. Holding it fixed within each instance allows LLMs and non-LLM baselines to be evaluated against identical alternatives. Each prediction instance is anchored at 14:00, providing a common temporal cut that preserves same-day mobility history while leaving an afternoon window in which to observe the next stay. The target is the first subsequent stay at a different location that begins within this window and lasts for at least ten minutes.

Each instance contains the individual's mobility history from preceding days, stays completed on the target day before 14:00, and the location occupied at the prediction cut. Under the attribute condition, age band, gender, occupation, and income level are added to the same mobility context. Locations are represented by a stable identifier and semantic category, providing both a consistent reference and interpretable information about location function. Each candidate is presented with its identifier, semantic category, and, under the default specification, distance from the current location. Candidate order is independently randomised for each instance to prevent presentation position from systematically indicating the target, given documented position sensitivity in LLM multiple-choice tasks (Zheng et al., 2024). Each request used fixed system and user-message templates. The system message required a JSON ranking of all 100 candidate identifiers, each used exactly once and drawn only from the supplied candidates. In the attribute condition a three-line block containing age band, gender, occupation and income level was inserted immediately before the current-location block, and all other content was identical between the paired conditions. Full templates, two de-identified paired examples and a line-by-line difference are provided in Supplementary Note.

Monday 18 March is used as the target day, with each individual contributing at most one prediction instance. Among the 5,000 sampled individuals, 1,979 had a qualifying target satisfying the temporal, location-change, and dwell-time criteria. Of these, 1,934 had mobility observations for all six preceding days. Individuals without a qualifying location change during the prediction window do not enter the benchmark. The implications of this eligibility condition are discussed in Section 4.3. Limiting the benchmark to one instance per individual also avoids repeated observations from the same person in paired model comparisons. The attribute experiments use a seeded random subsample of 1,000 benchmark instances selected before model evaluation. The remaining instances are reserved for the prompt-intervention experiments, ensuring that the two experiment sets contain no shared individuals. Automated checks confirmed that every candidate set contains exactly 100 distinct locations, exactly one target, and no negative candidate corresponding to the current location. Target position was randomised independently across candidate lists, with a mean position of 50.2, close to the expected midpoint of 50.5. The observed continuation had previously been visited by the same individual in 54.1% of the full benchmark and 55.3% of the 1,000-instance evaluation sample.

Two candidate-generation protocols are used to examine the sensitivity of reported prediction performance to benchmark construction. Under the proximity protocol, the 99 negative candidates are selected from locations nearest to the individual's current position. Because the observed continuation is inserted irrespective of its distance, it is the farthest of the 100 candidates in 61.7% of proximity lists against 0.2% of popularity lists. Under the popularity protocol, negatives are selected according to location visit frequency during the reference period preceding the target day. Individuals, mobility histories, and observed targets are held fixed across protocols. The resulting candidate sets nevertheless differ substantially in their spatial distribution, visit frequency, personal familiarity, and semantic composition. Table 1 summarises these differences in candidate-set composition.

**Table 1.** Candidate-set composition under the two negative-sampling protocols (n = 1,979 lists).

| Property | Proximity candidates | Popularity candidates |
|---|---|---|
| Distance from current location (km) | 0.30 | 17.82 |
| Distance of the true continuation (km) | 0.71 | 0.71 |
| Reference-period visit count | 1.0 | 360.5 |
| Share already in the person’s history (%) | 8.29 | 3.73 |
| Distinct categories per list | 39.5 | 10.7 |
| Distinct communities per list | 3.1 | 74.8 |

*Note. Values are medians for distance from the current location, distance of the true continuation, and reference-period visit count; pooled candidate-level percentages for the share already present in the person’s history; and means across candidate lists for the numbers of distinct categories and communities per list.*

The candidate universe is defined from distinct locations identified in the observed mobility records. Stays matched within 50 m to an entry in the Shenzhen POI catalogue are represented by their POI identifier, while unmatched stays are represented using a coordinate-grid identifier. Because the universe was assembled from the full observation week, it can contain locations first observed after the prediction cut. For 6.7% of the benchmark the target is a grid-identified location of this kind, so its presence in the universe depends on information observed later in the study window. Excluding these instances changes each paired attribute effect by no more than 0.11 percentage points (Section 3.2), and they are retained in the primary benchmark.

## 2.3 Experimental design

The primary analysis uses a paired design at the prediction-instance level. For each instance, two prompts are constructed that differ only in the inclusion of the sociodemographic attribute block, with the candidate set, mobility history, model, decoding configuration, and all other prompt content held fixed. The primary quantity of interest is the paired change in top-1 accuracy following attribute inclusion, with top-5 accuracy and mean reciprocal rank (MRR) reported as secondary metrics. McNemar's test is used for paired comparisons of binary top-1 and top-5 outcomes (McNemar, 1947), and percentile bootstrap confidence intervals (Efron & Tibshirani, 1993) for paired metric differences are estimated from 10,000 individual-level resamples.

An equivalence analysis is used to assess whether any change in top-1 accuracy is small enough to be considered practically negligible. The equivalence region is defined as ±2 percentage points. This margin was introduced as a practical threshold for an improvement considered insufficient to offset the additional prompt information and the potential stereotyping concerns associated with demographic attributes (Wu & Wang, 2024). Following the confidence-interval formulation of equivalence testing (Lakens, 2017), equivalence is concluded when the 90% percentile bootstrap confidence interval for the paired top-1 difference falls entirely within the interval [−0.02, 0.02]. Because the margin is absolute, it corresponds to a large relative change when baseline accuracy is low: at the $K = 0$ baseline of 5.6%, 2 percentage points is about 36% of baseline accuracy. The margin was also specified after the primary results had been examined (Section 4.3), so equivalence results are read as post hoc bounds on plausible effect size. For hypothesis families containing multiple related tests, such as the four attribute-recovery tests in Section 3.5, p values are adjusted using the Benjamini and Hochberg (1995) procedure.

The amount of prior mobility history is varied to examine whether the predictive contribution of attributes changes as behavioural information becomes more limited. The same prediction instances are evaluated with prior-day history truncated to the most recent $K \in \{0, 1, 2, 6\}$ days. Truncation is applied by calendar day. The rendered-history limit is set to 200 stays so that the intended history conditions remain distinct; a 24-stay limit would retain approximately 2.5 days at the observed stay density and truncate the longer conditions. At $K = 0$, all prior-day history is removed, while stays observed on the target day before the prediction cut and the current location remain available. This condition therefore evaluates prediction without prior-day history but still retains same-day behavioural context. A separate minimal-history condition, denoted $K = -1$, removes the target-day stay history in addition to all prior-day history. The available mobility information is then limited to the current location, with the candidate set supplied in both prompt conditions and sociodemographic attributes supplied only in the attribute condition. This condition is evaluated on a fixed subset of 500 individuals and is analysed separately from the history-length sequence because it changes the available behavioural context more substantially than truncating prior days alone. Its purpose is to assess whether the limited contribution of attributes under the main conditions can be explained by behavioural information already encoding individual-specific regularities.

The primary LLM is GPT-5.5, queried through a Chat Completions endpoint between July and August 2026 and JSON-constrained output. Each request and response is cached using a content hash, and the model identifier returned by the endpoint is recorded for every call to support auditing and replication. Selected experiments are repeated with two additional models using fixed nested subsets of the main evaluation sample. GPT-5, an earlier generation from the same provider, tests whether the principal patterns are specific to one model checkpoint. Claude Opus 4.6, developed independently by a different provider, tests whether they are specific to one provider's training pipeline. Both replications use identical prompts, candidate sets and scoring code; only the model identifier changes. Responses are parsed from JSON-constrained output and scored on the first occurrence of each identifier. A response is literally complete if it returns exactly 100 identifiers without repetition, and coverage-complete if every candidate receives a rank once duplicates are discarded. At $K = 6$ these rates are 61.2% and 73.6%.

Non-LLM baselines are evaluated on the same candidate sets and are classified according to whether their inputs are also available in the LLM prompt. Personal visit frequency, frequency followed by recency, first-order Markov transition probability, and distance from the current location use only information contained in the corresponding prediction instance. These matched-information baselines represent simple behavioural mechanisms based on recurrent mobility and spatial proximity. Reference-period popularity is treated separately because it ranks candidates using aggregate visit frequencies computed across the reference population, information that is not supplied to the LLM.

The attribute ablation is repeated on a supervised candidate reranker. It scores each candidate from twenty candidate-level features, is trained with person-level out-of-fold splits, and averages five seeds. All features use information available at the prediction cut, and the candidate distance annotation is excluded. Three arms differ only in the attribute block, which is absent, correct, or drawn from another person. Because attributes are constant across a person's 100 candidates they can act only through interactions, so fifteen explicit attribute by candidate product terms are included. The reranker is trained directly on this benchmark, and it is used as a second predictor with a different inductive bias.

# 3. Results

## 3.1 Behavioural history supplies the usable predictive signal

Person-specific mobility history provides a clear and usable predictive signal. In the no-attribute condition, top-1 accuracy rises monotonically from 0.056 at K = 0 to 0.185 at K = 6, an increase of 12.9 percentage points (Table 2; Figure 2a). Because the same prediction instances are evaluated across history conditions, this confirms that the benchmark is sensitive to additional person-specific mobility information, and it provides the reference against which the contribution of sociodemographic attributes is assessed in Section 3.2.

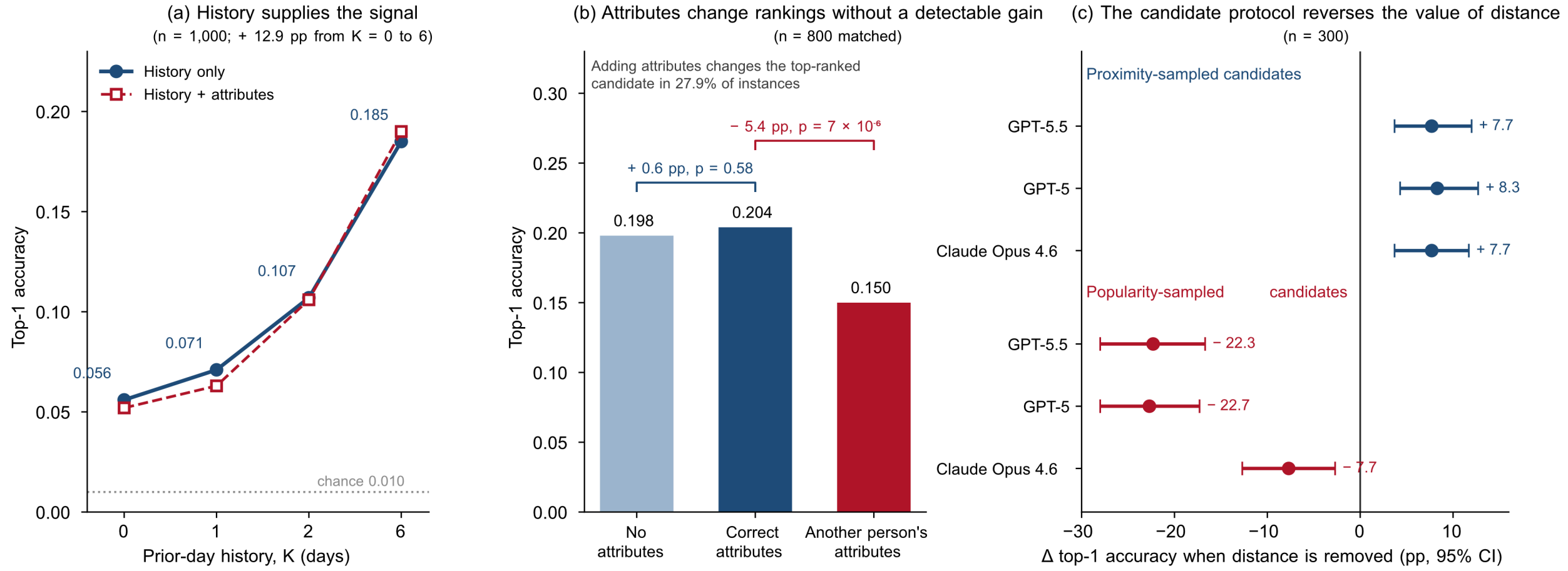

***Figure 2. Overview of the main findings.*** *(a) Top-1 accuracy by prior-day history, with and without attributes (n = 1,000). (b) Model response versus predictive benefit on the matched 800-individual subset: correct attributes leave accuracy essentially unchanged relative to no attributes, whereas attributes taken from another person reduce it. (c) Change in top-1 accuracy when the candidate distance annotation is deleted, under proximity and popularity sampling, for three LLMs (n = 300; 95% bootstrap confidence intervals).*

Next-location predictability differs sharply between revisited and previously unseen destinations. Among the 1,000 evaluation instances, 55.3% of observed continuations are locations that the individual has visited before, while 44.7% are first-time destinations. Prediction is extremely difficult in the latter group. The LLM achieves an Acc@1 of only 0.004 on first-time continuations, below the 0.010 expected from uniform random ranking over 100 candidates, and personal-history heuristics record no top-1 hits because the target is absent from the observed visitation history. For any method restricted to previously visited locations, the 55.3% revisit share therefore places a natural ceiling on benchmark-level top-1 accuracy. Furthermore, most of the usable predictive signal is instead concentrated in revisitation. Among revisit cases, ranking candidates by the individual's own visit frequency achieves an Acc@1 of 0.553, compared with 0.331 for the LLM. A simple rule combining personal frequency and recency reaches 0.330 across the full benchmark, 14.5 percentage points above the history-only LLM. These results indicate that much of the benchmark's predictability comes from person-specific repetition.

### 3.2 Demographic attributes change predictions without a detectable accuracy gain

In contrast to behavioural history, adding sociodemographic attributes produces no detectable improvement in top-1 accuracy at any level of prior mobility history. The paired differences are −0.4, −0.8, −0.1, and +0.5 percentage points at K = 0, 1, 2, and 6, respectively, with all McNemar p values ≥ 0.332 (Table 2). Under the ±2-percentage-point equivalence region introduced in Section 2.3, the 90% bootstrap confidence interval lies entirely within the equivalence bounds at K = 0, 2, and 6. At K = 1, the lower bound reaches the negative equivalence margin, so equivalence is not concluded. The remaining uncertainty in this condition concerns a possible reduction in accuracy. However, the absence of an accuracy gain does not imply that sociodemographic attributes have no effect on model predictions. The following analyses therefore distinguish predictive benefit from model responsiveness. First, the cold-start setting tests whether an attribute benefit emerges when behavioural history is unavailable. Second, matched comparisons among missing, correctly matched and mismatched attributes assess whether changing attribute content alters predictions even when it does not improve overall accuracy. The full set of diagnostic analyses is summarised in Figure 3.

**Table 2.** Marginal contribution of sociodemographic attributes by prior-history length (n = 1,000).

| K | No attr. | +Attr. | Δ | 95% CI | 90% CI | *b*:*c* | McN. *p* | Equivalent at ±.02 |
|---|---|---|---|---|---|---|---|---|
| 0 days | 0.056 | 0.052 | −0.004 | [−0.017, +0.009] | [−0.015, +0.007] | 20:24 | 0.652 | Yes |
| 1 day | 0.071 | 0.063 | −0.008 | [−0.022, +0.006] | [−0.020, +0.004] | 22:30 | 0.332 | No (harm side) |
| 2 days | 0.107 | 0.106 | −0.001 | [−0.017, +0.015] | [−0.015, +0.013] | 34:35 | 1.000 | Yes |
| 6 days | 0.185 | 0.190 | +0.005 | [−0.010, +0.020] | [−0.007, +0.018] | 32:27 | 0.603 | Yes |

*Note. Δ is top-1 accuracy with attributes minus accuracy without attributes. b denotes instances correct only with attributes and c instances correct only without attributes. Confidence intervals are obtained from 10,000 paired bootstrap resamples over individuals. Equivalence is concluded when the 90% bootstrap confidence interval lies entirely within [−0.02, 0.02].*

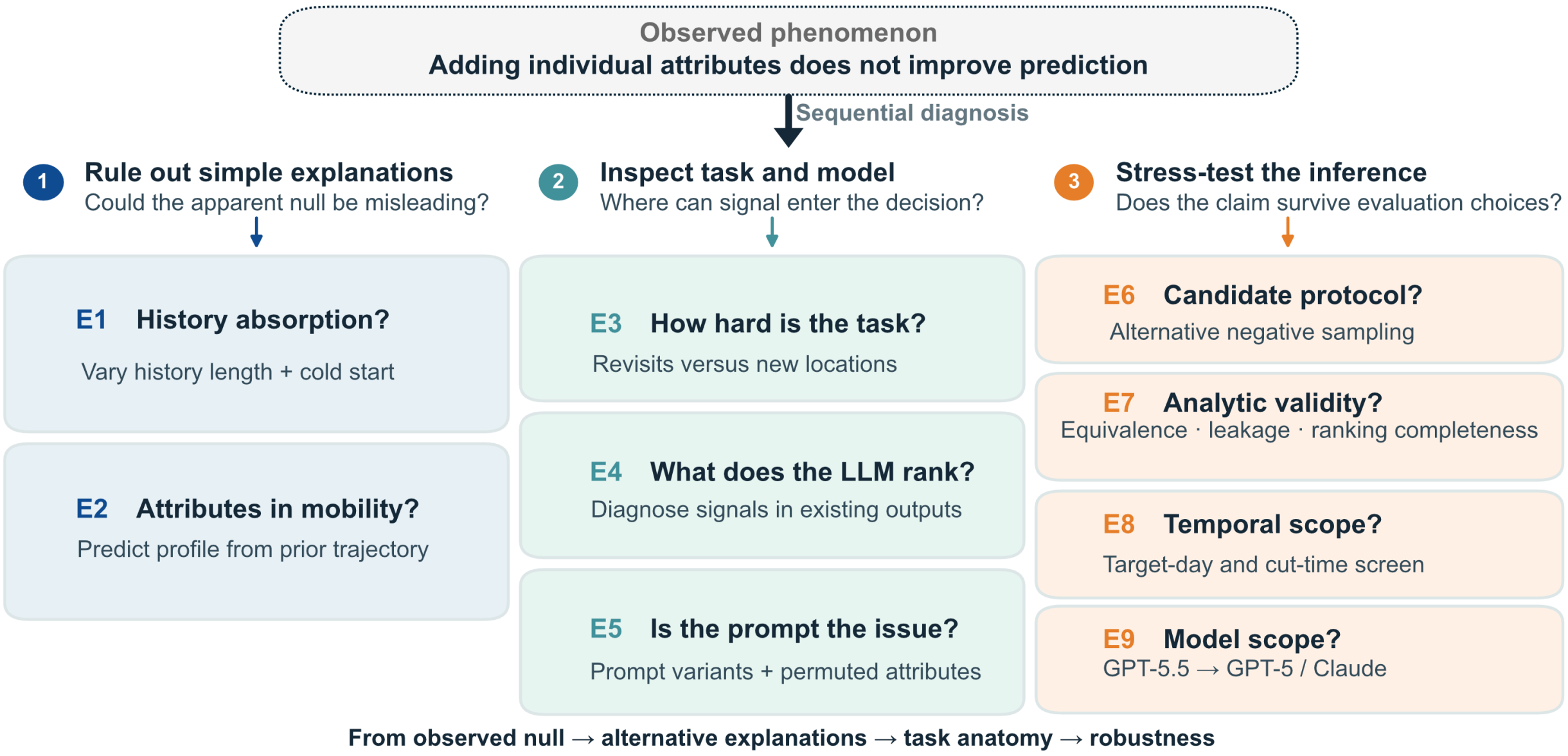

***Figure 3.*** *Diagnostic roadmap for the attribute null. The analyses examine behavioural redundancy, task structure, model response, benchmark construction, and evaluation choices.*

A first possible explanation for the absence of an attribute gain is that mobility history already captures much of the predictive information associated with sociodemographic characteristics. If so, sociodemographic attributes should become more informative when observed mobility history is removed. At $K = -1$, where both prior-day and same-day stay histories are withheld, top-1 accuracy is 0.026 without attributes and 0.024 with attributes. The paired difference is −0.002, with a 95% bootstrap confidence interval of [−0.010, +0.006] across 500 individuals (McNemar $p = 1.000$). This interval lies entirely within the ±2-percentage-point equivalence region. Removing observed mobility history therefore does not reveal a predictive benefit from sociodemographic attributes, providing little support for the explanation that their contribution is masked by information already contained in mobility history. However, predictive accuracy in this cold-start condition remains only slightly above chance, limiting how strongly this analysis can rule out redundancy between the two information sources.

The model is clearly responsive to attribute information, but this responsiveness does not translate into higher predictive accuracy. Adding the attribute block changes the top-ranked candidate in 27.9% of instances, and the median Kendall's $\tau$ between rankings with and without attributes is 0.683. The permutation experiment makes this distinction clearer. Among the same individuals, replacing each person's attributes with another person's demographic profile reduces top-1 accuracy from 0.204 to 0.150, a paired decline of 5.4 percentage points with a 95% confidence interval of [−0.078, −0.031] and a 24:67 discordant split ($p = 7 \times 10^{-6}$). Relative to the no-attribute condition, permuted attributes reduce accuracy by 4.8 percentage points ($p = 6 \times 10^{-5}$), whereas correct attributes increase it by only 0.6 percentage points ($p = 0.576$). The model therefore does not ignore the attribute block: attribute content influences its rankings, but correctly matched sociodemographic information provides no detectable predictive gain (Figure 2b). The result is also robust to two implementation choices. Alternative scoring of truncated rankings changes Acc@5 and MRR by less than 0.1 percentage points, while excluding the 67 instances affected by candidate-universe look-ahead changes the paired attribute effect by no more than 0.11 percentage points.

### 3.3 The absence of a gain extends across predictors and prediction times

The absence of a detectable gain is not specific to the primary model. Replication with two additional models produces the same qualitative result. At $K = 6$ on the same 300 individuals, the paired attribute effect is +0.010 for GPT-5, with a 95% confidence interval of [−0.023, +0.043] (p = 0.690), and +0.007 for Claude Opus 4.6, with a 95% confidence interval of [−0.017, +0.030] (p = 0.791). The corresponding estimate for GPT-5.5 is −0.017, with a 95% confidence interval of [−0.047, +0.013] (p = 0.383). None of the three models therefore shows evidence that adding sociodemographic attributes improves top-1 accuracy. These replications broaden the result beyond the primary model, although their confidence intervals are not sufficiently narrow to establish equivalence within the ±2-percentage-point margin. The upper bounds remain +4.3 percentage points for GPT-5 and +3.0 percentage points for Claude Opus 4.6, so the conclusion is an absence of detectable improvement from attribute information. Figure 4 summarises the marginal attribute effects across the main and diagnostic conditions alongside the gains associated with additional mobility history.

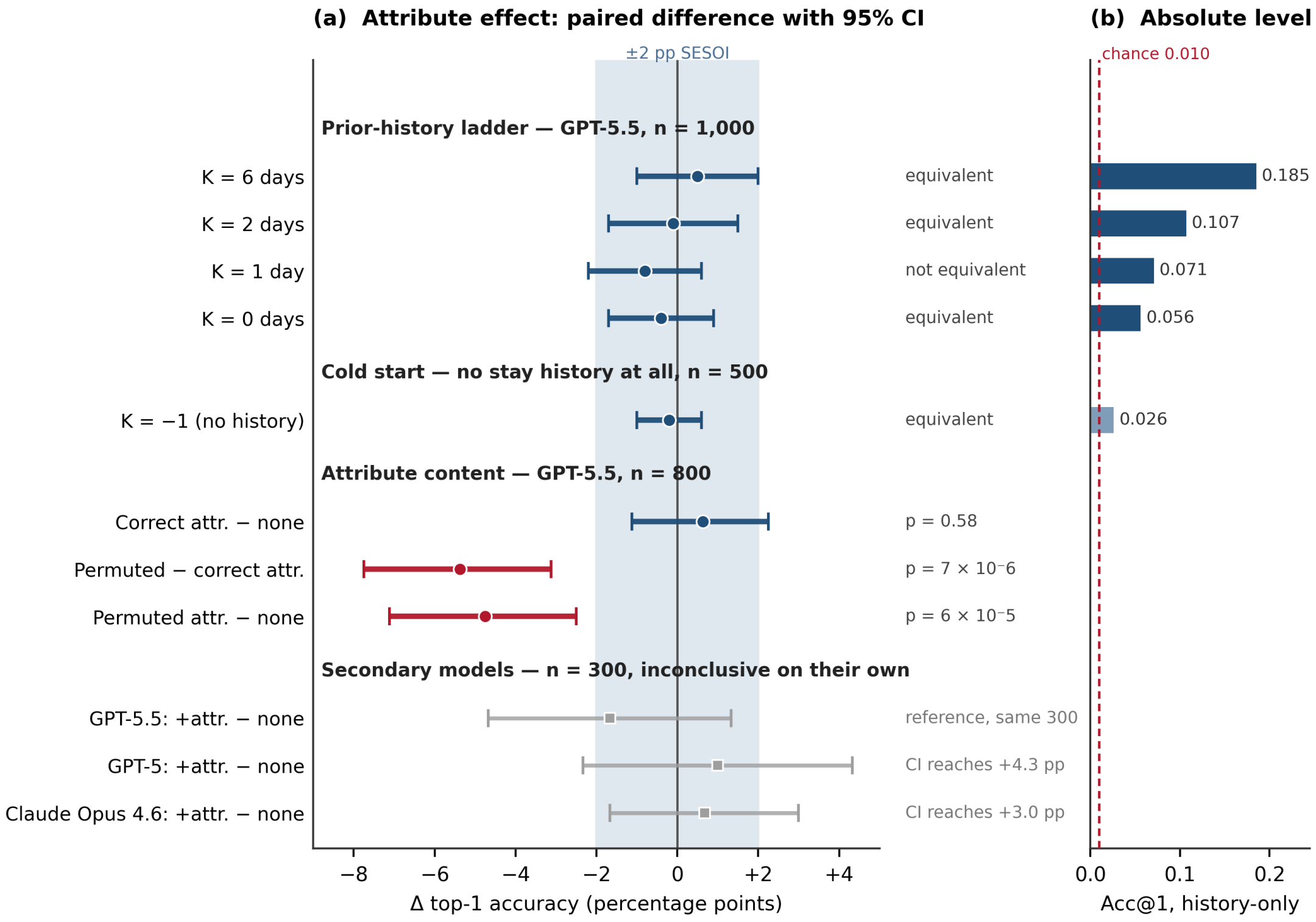

***Figure 4. Marginal contribution of sociodemographic attributes to top-1 accuracy.*** *(a) Paired differences with 95% person-level bootstrap confidence intervals; the shaded region is the ±2-percentage-point equivalence region introduced in Section 2.3. Grey squares are the two secondary models, GPT-5 and Claude Opus 4.6. (b) Absolute top-1 accuracy without attributes. The minimal-history condition is shown separately because it removes substantially more behavioural information and has correspondingly lower baseline accuracy.*

Temporal sensitivity analyses show similarly small attribute effects across alternative days and prediction times (Table 3). The benchmark was reconstructed for an additional weekday, a weekend day, and cut times of 08:00 and 18:00 on the primary target day. Each condition includes 200 individuals drawn from the main evaluation sample and uses all mobility history available before the corresponding target day. Paired differences range from −2.0 to +1.0 percentage points, with none reaching statistical significance. Because the confidence intervals remain wide enough to include effects beyond the ±2-percentage-point equivalence margin, these analyses are interpreted as sensitivity checks.

**Table 3.** Temporal sensitivity of the attribute effect using all available prior history (n = 200 per sensitivity cell).

| Cell | Prior days | No attr. | +Attr. | Δ | 95% CI | McN. *p* |
|---|---|---|---|---|---|---|
| 2024-03-18, cut 14:00 (main, n = 1,000) | 6 | 0.185 | 0.190 | +0.005 | [−0.010, +0.020] | 0.603 |
| 2024-03-15 (Fri), cut 14:00 | 3 | 0.105 | 0.115 | +0.010 | [−0.020, +0.040] | 0.754 |
| 2024-03-17 (Sun), cut 14:00 | 5 | 0.070 | 0.075 | +0.005 | [−0.025, +0.040] | 1.000 |
| 2024-03-18, cut 08:00 | 6 | 0.160 | 0.145 | −0.015 | [−0.060, +0.030] | 0.678 |
| 2024-03-18, cut 18:00 | 6 | 0.180 | 0.160 | −0.020 | [−0.055, +0.015] | 0.388 |

*Note. Sensitivity cells use all mobility history available before the corresponding target day.*

The weak attribute effect also persists when the LLM is replaced by a predictor with a different inductive bias. This comparison addresses the possibility that the small effect observed for the LLM reflects properties of the model itself. The same attribute ablation is therefore repeated with a supervised reranker trained directly on the benchmark, using identical candidate sets and predictor information, with sociodemographic attributes entered as structured predictors. Adding the attribute block changes top-1 accuracy by −0.94 percentage points over the full sample (95% CI [−1.88, +0.02]). The null effect also persists in two diagnostic subsets (Table 4). Among exploration targets, where the destination has not appeared previously in the individual's trajectory, adding attributes changes Acc@1 by only −0.22 percentage points (95% CI [−1.57, +1.03]). Restricting the analysis to cases where the target is not the farthest candidate, thereby reducing the candidate-geometry cue examined in Section 3.4, produces a similarly small change of −0.58 points (95% CI [−1.75, +0.58]). Neither subset shows evidence that sociodemographic attributes improve prediction.

The same pattern extends to the broader ranking. Adding the correct attribute block reduces Acc@5 by 1.26 percentage points relative to the no-attribute condition (95% CI [−2.16, −0.36]), while permuted attributes produce a change of −0.94 points (95% CI [−1.90, +0.02]). The difference between correctly matched and permuted attributes is only -0.32 points (95% CI [−0.70, +1.36]), giving no detectable advantage to the correct demographic profile. This differs from the LLM, for which the same permutation reduces top-1 accuracy by 5.4 percentage points on the matched 800-individual subset. Thus, the limited predictive value of sociodemographic attributes extends to the supervised reranker, where correctly matched profiles provide no detectable advantage over permuted ones.

**Table 4.** Marginal contribution of the attribute block to the supervised reranker (n = 1,000).

| Arm | Acc@1 | Acc@5 | Revisit Acc@1 (n=553) | Exploration Acc@1 (n=447) | Target not farthest Acc@1 (n=377) |
|---|---|---|---|---|---|
| No attributes | 0.3478 | 0.5966 | 0.5168 | 0.1387 | 0.2255 |
| Correct attributes | 0.3384 | 0.5840 | 0.5016 | 0.1365 | 0.2196 |
| Permuted attributes | 0.3404 | 0.5872 | 0.5060 | 0.1356 | 0.2244 |

Note. *Five seeds with person-level out-of-fold splits. Paired 95% bootstrap intervals for correct minus no attributes are −0.94 pp [−1.88, +0.02] overall, −1.52 pp [−2.89, −0.18] on revisits, −0.22 pp [−1.57, +1.03] on exploration, and −0.58 pp [−1.75, +0.58] on lists where the target is not the farthest. Intervals are computed over the five-seed mean; McNemar's test uses a single seed and is reported only for the full-sample Acc@5 contrast (p = 0.022).*

## 3.4 Candidate construction changes judgements about information and models

Candidate construction strongly shapes the measured value of prompt information by changing the alternatives against which the observed target is ranked. On the same 300 individuals, GPT-5.5 attains Acc@1 = 0.217 under proximity sampling but 0.643 under popularity sampling, despite receiving the same mobility history and prediction task. Distance provides a clear mechanism for this difference. The LLM ranking is more closely aligned with candidate distance than with personal visit frequency, with Spearman correlations of 0.693 and 0.348, respectively. Under proximity sampling, negative candidates are selected near the current location while the observed target is inserted without a distance constraint, making greater distance disproportionately indicative of the target. The same distance information therefore has very different predictive value under the two protocols. Furthermore, removing the distance annotation confirms this dependence. Under proximity sampling, deleting distance increases GPT-5.5 top-1 accuracy from 0.217 to 0.293, whereas under popularity sampling it reduces accuracy from 0.643 to 0.420 (Table 5; Figure 5a). The same sign reversal appears for GPT-5 and Claude Opus 4.6, although the decline under popularity sampling is smaller for Claude. The effect of a fixed prompt component therefore depends strongly on the alternatives against which the target is evaluated.

**Table 5.** Effect of removing the candidate distance annotation, by predictor and candidate protocol.

| Predictor | Developer | Distance present → removed | Δ | 95% CI | p |
|---|---|---|---|---|---|
| Proximity-sampled negatives | | | | | |
| GPT-5.5 | OpenAI | 0.217 → 0.293 | +7.7 pp | [+3.7, +12.0] | $6.1 \times 10^{-4}$ |
| GPT-5 | OpenAI | 0.193 → 0.277 | +8.3 pp | [+4.3, +12.7] | $1.7 \times 10^{-4}$ |
| Claude Opus 4.6 | Anthropic | 0.207 → 0.283 | +7.7 pp | [+3.7, +11.7] | $4.3 \times 10^{-4}$ |
| Popularity-sampled negatives | | | | | |
| GPT-5.5 | OpenAI | 0.643 → 0.420 | −22.3 pp | [−28.0, −16.7] | $6.2 \times 10^{-13}$ |
| GPT-5 | OpenAI | 0.570 → 0.343 | −22.7 pp | [−28.0, −17.3] | $1.7 \times 10^{-15}$ |
| Claude Opus 4.6 | Anthropic | 0.563 → 0.487 | −7.7 pp | [−12.7, −2.7] | $3.2 \times 10^{-3}$ |

*Note. Top-1 accuracy, n = 300 on the same individuals in all six cells, parse rate 1.000 throughout. Rows one to three use proximity-sampled negatives and rows four to six popularity-sampled negatives.*

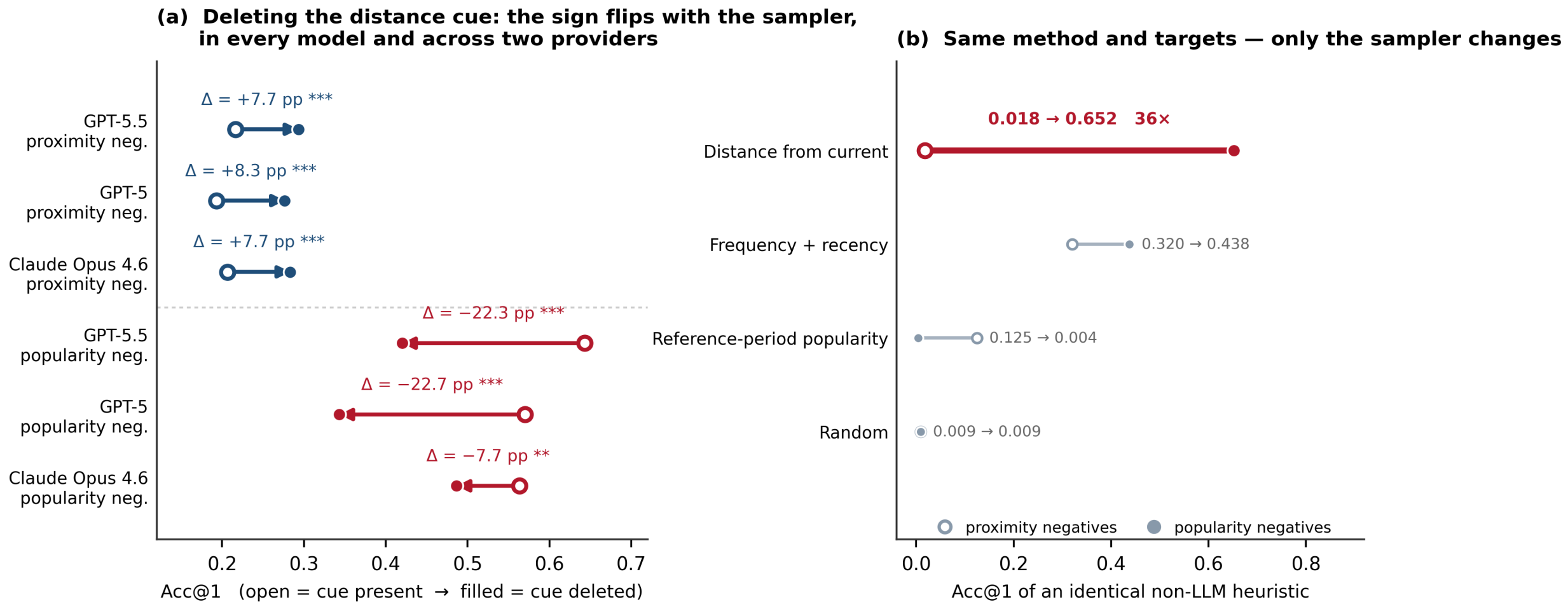

***Figure 5. The candidate-generation protocol governs the value of a distance cue.*** *(a) Deleting the distance annotation, with every other token unchanged, raises accuracy under proximity-sampled negatives and lowers it under popularity-sampled negatives; open circles are cue-present, filled cue-deleted. The reversal holds in all three models and across two providers (p < 0.05 in every cell; Table 5), though it is markedly weaker for Claude*

*Opus 4.6 under the popularity protocol. (b) On the full benchmark, changing only the candidate protocol shifts nearest-distance accuracy from 0.018 to 0.652, while random ranking is unchanged.*

The influence of candidate construction is evident even with distance-only heuristics. Across the full benchmark, a farthest-first rule attains Acc@1 = 0.616 under proximity sampling but only 0.002 under popularity sampling, while a nearest-first rule changes in the opposite direction from 0.018 to 0.652. Random ranking remains essentially unchanged at 0.009 under both protocols (Figure 5b). Because these heuristics use no information about the individual, mobility history, or location semantics, their large performance shifts arise directly from how the candidate pool is constructed. Under proximity sampling, the farthest-first rule even exceeds every predictive model evaluated in this study. This structure also bounds how far their results generalise to candidate sets built for deployment, because the attribute experiments in Sections 3.2 and 3.3 used the proximity protocol.

Candidate construction also changes the relative ordering of competing predictors. On the same 300 instances, the frequency-and-recency heuristic outperforms the LLM under proximity sampling, with Acc@1 of 0.320 versus 0.217. Under popularity sampling, the ordering reverses, with the LLM reaching 0.643 compared with 0.438 for the heuristic. The choice of candidate-generation protocol therefore affects both the level of reported accuracy and conclusions about which method performs better. When negative candidates are sampled according to variables related to destination choice, those variables can themselves become informative of target status, making benchmark performance inseparable from the construction of the candidate pool.

Within the primary proximity benchmark, candidate-list geometry also explains a substantial part of the performance differences between methods. When the target is not the farthest candidate, which occurs in 37.7% of instances, LLM accuracy rises from 0.185 overall to 0.276, narrowing its gap with the frequency-and-recency heuristic to 3.7 percentage points and placing it above the supervised reranker. When the target is the farthest candidate, by contrast, LLM accuracy falls to 0.130 while reranker accuracy rises to 0.422 (Table 6). The two models therefore respond in opposite directions to the same candidate structure, whereas the frequency-and-recency heuristic remains comparatively stable. This indicates that aggregate model comparisons partly reflect candidate construction. The candidate effect is also important for interpreting apparent success on exploratory destinations. The supervised reranker attains 13.9% accuracy across all first-time targets, but much of this performance disappears when the strongest geometric cue is removed. Among the 142 exploration instances in which the target is not the farthest candidate, accuracy falls to 0.023 for the reranker and 0.007 for the LLM, while the frequency-and-recency heuristic remains at 0.000; all are close to the 0.010 random-ranking benchmark. Thus, these results show that benchmark predictability is concentrated in revisitation, while apparent success on exploratory destinations is strongly influenced by candidate geometry.

**Table 6.** Method performance by candidate-list structure (n = 1,000, proximity protocol).

| Subset | n | Share | Supervised reranker | Frequency + recency | LLM |
|---|---|---|---|---|---|
| All lists | 1,000 | 100% | 0.348 | 0.330 | 0.185 |
| Target is the farthest candidate | 623 | 62.3% | 0.422 | 0.340 | 0.130 |
| Target is not the farthest | 377 | 37.7% | 0.225 | 0.313 | 0.276 |
| Exploration and not farthest | 142 | 14.2% | 0.023 | 0.000 | 0.007 |

*Note. Top-1 accuracy. Chance is 0.010 throughout.*

### 3.5 Sociodemographic information is partly recoverable from mobility trajectories

One possible explanation for the weak predictive value of sociodemographic attributes is that these characteristics are only weakly related to mobility behaviour in the study population. To examine this possibility, the prediction direction is reversed by testing whether mobility behaviour can recover the same sociodemographic characteristics used in the next-location experiments. Eighteen features derived from stay sequences before the 14:00 prediction cut capture spatial range, activity intensity, temporal regularity, revisitation patterns and the diversity of visited locations. Gradient-boosted classifiers use the same feature set to predict each of the four sociodemographic characteristics separately. Predictive performance varies substantially across attributes. Above-median income is recovered with an out-of-fold AUC of 0.708 (Table 7), whereas gender, age group and white-collar status are predicted only weakly, with AUCs ranging from 0.545 to 0.559. All trajectory features are derived from observations available before the prediction cut, ensuring that the two predictive directions use temporally comparable information.

**Table 7.** Recovery of sociodemographic attributes from pre-cut trajectory features (n = 5,000).

| Attribute | AUC | 95% CI | Permutation null (SD) | Distance from null | *p* |
|---|---|---|---|---|---|
| Income above median | **0.708** | [0.695, 0.721] | 0.502 (0.010) | 20.6 SD | < 0.05 |
| Male | 0.559 | [0.544, 0.573] | 0.499 (0.010) | 6.0 SD | < 0.05 |
| Age ≥ 40 | 0.555 | [0.539, 0.570] | 0.504 (0.010) | 5.1 SD | < 0.05 |
| White-collar | 0.545 | [0.531, 0.559] | 0.500 (0.007) | 6.4 SD | < 0.05 |

***Note.*** *AUC is estimated using five-fold stratified cross-validation repeated three times. Permutation tests use 40 label shuffles, giving a minimum attainable p = 0.024 (BH-adjusted). Benjamini–Hochberg adjustment is applied across the four attribute tests.*

These results confirm that mobility trajectories contain sociodemographic signal, most clearly for income, although the remaining attributes are only weakly recoverable. The predictive value of this information nevertheless depends on the task in which it is used. Recovering a demographic attribute is a binary classification problem evaluated by AUC, whereas next-location prediction requires ranking 100 candidate destinations conditional on the individual's own mobility history. Performance in one task therefore does not directly translate into predictive gain in the other. More importantly, the reverse-prediction results show that explicit demographic labels are not the only source of sociodemographic information available to a mobility model, because some of that information is already encoded in observed trajectories. This has direct privacy implications, discussed further in Section 4.1.

---

## 4. Discussion

### 4.1 Implications for persona-conditioned mobility modelling

The results distinguish the role of sociodemographic information in representing population-level mobility heterogeneity from its value for individual next-location prediction. Sociodemographic characteristics are well-established correlates of mobility behaviour and can help reproduce population heterogeneity in mobility simulation (Tzachristas et al., 2026; Li et al., 2024). Related work also shows that accumulated mobility traces can reveal socioeconomic characteristics (Blumenstock et al., 2015; Uğurel et al., 2026), a pattern reproduced by the attribute-recovery analysis in this study. Yet neither population-level association nor successful attribute recovery establishes that these characteristics improve prediction of a particular individual's next destination. The paired experiments show that sociodemographic information can be reflected in mobility behaviour without providing comparable incremental value once that individual's mobility history is already observed.

The recurrent structure of human mobility offers one explanation for this pattern. Individuals repeatedly return to previously visited locations while intermittently exploring new ones (Pappalardo et al., 2015), producing persistent spatial and temporal regularities across populations and cities (Schläpfer et al., 2021). Exploration events set the limit on how far any predictor can go (Song et al., 2010; Cuttone et al., 2018). For short-horizon prediction, an individual's own visitation history provides direct evidence about likely destinations, particularly when the next trip returns to a familiar location. Sociodemographic attributes offer only broader information about typical behaviour among similar people. This difference is reflected in the benchmark, where most of the usable predictive signal comes from revisitation, while first-time destinations remain difficult to predict.

The findings also sharpen the distinction between LLMs' responsiveness to demographic cues and their predictive usefulness. Wu and Wang (2024) show that demographic prompting can substantially shift predicted destinations while amplifying demographic stereotypes beyond the differences observed in real behaviour. The present study extends this concern to individual-level prediction using individual-level linked attributes and mobility histories, and we find that attributes alter model rankings without producing a measurable gain in accuracy. Demographic conditioning should therefore not be assumed to improve prediction simply because the model responds strongly to it. Matched comparisons of missing, correct and mismatched attributes help separate model responsiveness from predictive value, because mismatched profiles can reveal sensitivity to attribute content even when correct profiles do not improve accuracy. In mobility simulation, such attributes may still be useful for representing population heterogeneity, but their behavioural realism needs to be validated. Their predictive value may also differ in settings where individual routines provide less information and socioeconomic constraints more directly shape travel choices.

The results also have two implications for data privacy. First, the limited predictive value of sociodemographic attributes weakens the case for collecting or retaining such information solely to improve next-location prediction. Second, the reverse-prediction analysis shows that omitting these attributes does not remove the underlying privacy concern, because mobility trajectories themselves can reveal socioeconomic characteristics, particularly income. This extends previous evidence that mobility traces can expose sensitive information about individuals (Blumenstock et al., 2015). The information needed for a prediction task and the information that can be inferred from the data should therefore be considered separately when assessing privacy risk.

### 4.2 Candidate construction and comparability in next-location evaluation

Candidate construction can substantially shape measured prediction accuracy by changing the relationship between predictive cues and target status. The distance experiments in this study provide a clear example. Under proximity sampling, nearby locations are disproportionately selected as negatives, whereas under popularity sampling they are not, causing the same distance cue to impair prediction in one protocol and improve it in the other. The sign reversal observed across all three LLMs shows that the predictive value of a feature cannot be interpreted independently of how candidate alternatives are generated. This issue is well recognised in sampled recommender evaluation, where negative sampling can distort ranking metrics and relative model comparisons (Krichene & Rendle, 2020; Bae et al., 2024; Prakash et al., 2024). It is especially consequential for mobility prediction because variables commonly used to construct candidate sets, including distance and destination popularity, are themselves meaningful predictors of travel behaviour. The present results show how this interaction operates for LLM predictors. Because the models respond strongly to explicit distance annotations, spatial sampling determines whether the same distance field helps or harms, and the reversal appears in all three LLMs.

Model performance reported under different candidate-generation procedures should therefore not be compared directly as evidence of differences in predictive capability, consistent with broader concerns about reproducibility and baseline comparability in recommender-system evaluation (Ferrari Dacrema et al., 2019; Krichene & Rendle, 2020). At minimum, next-location benchmarks should report the candidate universe and negative-sampling procedure explicitly and evaluate proposed models and behavioural baselines on identical candidate sets. For LLM-based evaluation, the same principle extends to the information supplied in the prompt, an object whose components can contribute differently to model behaviour (Zheng et al., 2026). The attribute block in this study changed model rankings without improving accuracy, while the value of the distance field depended strongly on the candidate protocol. Meaningful comparison therefore requires both the model inputs and the construction of the alternatives being ranked to be specified and held comparable.

### 4.3 Limitations

The first limitation concerns the scope of the prediction setting. At the benchmark level, the primary analysis is defined by a single target day, a 14:00 prediction cut, and a four-hour horizon; the additional weekday, weekend, and cut-time analyses provide only exploratory evidence beyond this configuration. More fundamentally, the task is formulated as closed-set ranking over a fixed candidate list. The finding that sociodemographic attributes add little under this setting therefore does not imply that they would be uninformative in candidate generation or open-ended location prediction. Beyond the task formulation itself, the data cover only one city and one week, further limiting generalisation to mobility systems with different spatial structures, travel rhythms, or sociodemographic compositions.

The second limitation concerns the measurement of sociodemographic attributes and sample selection. The study uses four operator-derived categorical attributes, age band, gender, occupation, and income level, which capture common dimensions of sociodemographic profile but remain relatively coarse. The finding of limited incremental predictive value should therefore be interpreted in relation to this form of attribute representation; richer variables more directly linked to travel constraints and opportunities, such as household composition and vehicle access, may warrant separate examination. In addition, the analytical sample may overrepresent individuals with stable or consistently observed mobility patterns, limiting generalisation to populations with more irregular mobility or less complete observation.

The third limitation concerns the scope of the mechanisms and analyses examined. The distance experiments identify a reproducible behavioural response to adding or removing candidate-distance information, but they do not reveal how the models internally represent or use that cue. Similarly, the attribute-recovery analysis uses eighteen aggregate trajectory features and does not establish the maximum sociodemographic information recoverable from richer sequential representations. The two deliberately contrasting negative-sampling protocols show that candidate construction can substantially alter reported accuracy and relative method performance, but they do not characterise the broader space of candidate-sampling strategies. Finally, the ±2-percentage-point equivalence margin and the equivalence analysis were specified only after the primary results had been examined. They should therefore be interpreted as post hoc bounds on the plausible effect size. In addition, the cross-model replications have confidence intervals that still allow gains of 3 to 4 percentage points, and accuracy in the minimal-history condition is only 2.6%, so neither analysis fully excludes a modest attribute benefit or redundancy between history and attributes.

## 5. Conclusion

This study examined which personal information improves LLM-based next-location prediction and how its contribution can be verified. Behavioural history provided a clear predictive signal, with accuracy rising steadily as more prior-day history was supplied. By contrast, adding age, gender, occupation and income produced no detectable improvement in predictive accuracy in the primary paired experiments. The model nonetheless responded clearly to demographic content, as attributes altered its rankings and deliberately mismatched attributes reduced accuracy, ruling out the possibility that the attribute block was simply ignored. Removing stay history did not uncover a stronger attribute effect, suggesting that the null cannot be explained simply by mobility history absorbing the same information. A supervised reranker trained on the same benchmark reproduced the weak attribute effect, indicating that the result is not confined to a particular LLM-based ranking mechanism. Taken together, these findings distinguish model responsiveness to personal information from its incremental predictive value, and demographic association from predictive usefulness. Sociodemographic characteristics can be reflected in mobility behaviour and, particularly for income, recovered from accumulated trajectories, yet they add little information for predicting an individual's immediate next destination in this setting.

A separate finding concerns the evaluation setting itself. Candidate construction substantially changes measured accuracy by altering the relationship between predictive cues and the alternatives being ranked. The same distance information helps under one candidate protocol and harms under another, and changing the protocol can even reverse the relative performance of an LLM and a simple behavioural heuristic. Next-location accuracy should therefore be interpreted in relation to the predictor, the information supplied to it and the candidate-generation procedure. More broadly, the predictive value of personal information should be established through matched comparisons under clearly specified candidate-generation conditions. Such evidence is particularly important for sociodemographic attributes, whose inclusion entails additional privacy and data-governance considerations.

---